\documentclass[12pt]{elsarticle}

\usepackage{graphics}
\usepackage{amssymb}
\usepackage{amsmath}
\usepackage{mathtext}
\usepackage{nicefrac}

\makeatletter
\def\ps@pprintTitle{%
 \let\@oddhead\@empty
 \let\@evenhead\@empty
 \def\@oddfoot{\centerline{\thepage}}%
 \let\@evenfoot\@oddfoot}
\makeatother

\begin{document}

\begin{frontmatter}

\title{A possibility of interpretation of the cosmic ray knee near 10 TV as a contribution of a single close source}

\author[sinp]{D. Karmanov}
\author[sinp]{I. Kovalev}
\author[sinp]{I. Kudryashov\corref{cor}}
\cortext[cor]{Corresponding author}
\ead{ilya.kudryashov.85@gmail.com}
\author[sinp]{A. Kurganov}
\author[mex]{V. Latonov}
\author[sinp]{A. Panov}
\author[sinp]{D. Podorozhnyy}
\author[sinp]{A. Turundaevskiy}

\address[sinp]{Skobeltsyn Institute of Nuclear Physics, Moscow State University, 1(2), Leninskie Gory, GSP-1, Moscow, 119991, Russia}
\address[mex]{Department of Mechanics and Mathematics, Moscow State University, 1, Leninskie Gory, GSP-1, Moscow, 119991, Russia}

\begin{abstract}


This article presents a description of a cosmic rays diffusive propagation model of a close point-like flash lamp like source and an approximation of experimentally observed spectral irregularity with this model. We show that this spectral irregularity can be explained using the presented model and provide the most probable characteristics of such a source as well as several observed and identified sources which can be candidates for this role.

\end{abstract}


\end{frontmatter}

\parindent=0.5 cm


\section{Introduction}


Results (data) of a combination of cosmic ray (CR) physics experiments \citep{panov,hawc,cream} point to a CR inclination change near magnetic rigidity of 10 TV (we'll refer to it as the ``small knee'' from now on). The direct measurement NUCLEON space experiment allows to measure the structure of the ``small knee'' for each of the abundant primary CR components separately, which is important to understand its nature.


Such an irregularity in the normally regular CR spectra can be explained by several possible reasons: CR acceleration features (for example, an acceleration limit in specific types of the supernovae remnants), CR propagation features (for example, anomal diffusion \citep{lagutin}) or a contribution to the CR flow of a single close source \citep{chkink}. A significant abruptness of the inclination change in terms of magnetic rigidity \citep{hawc} is an indirect indication to the ``small knee'' being determined by an acceleration limit of the CR in a single close source, like a supernova remnant. If this inclination change was contributed to by several sources, then it would be difficult to expect an almost exactly coinciding acceleration limit of them all, which can be the only explanation of the abruptness of the knee. Therefore, in this article only a single close source is explored as an explanation to the ``small knee'' phenomenon.


To test this hypothesis, authors created a mathematical model of CR flow from a close source. The model is based on solving the CR propagation problem in diffusion approximation. Free parameters of the model were approximated to fit experimental data, and regions of acceptable and most probable values of these parameters were determined.

\section{Mathematical model}

\begin{figure}[!t]
\begin{center}
\includegraphics*[width=0.9\textwidth]{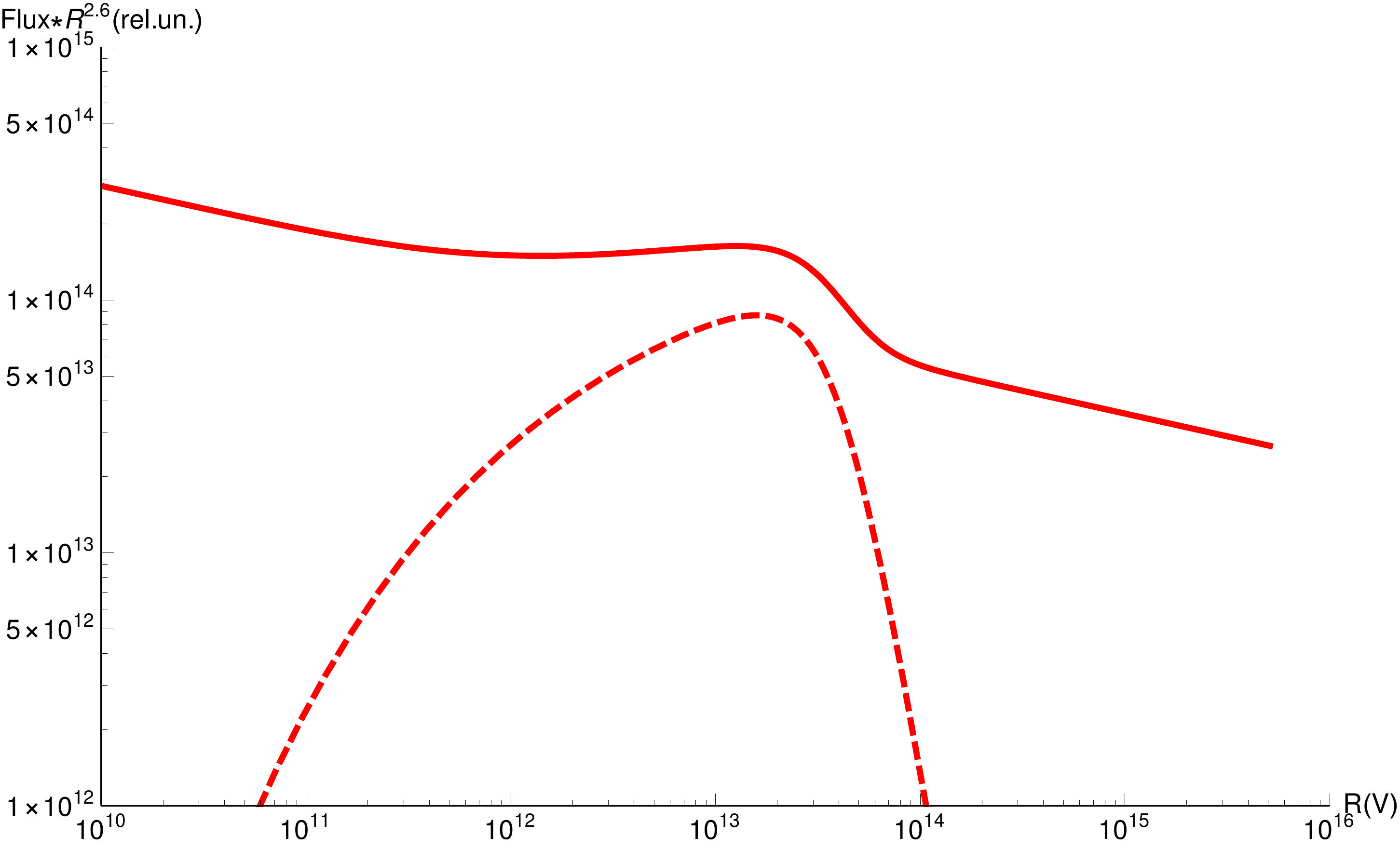}
\end{center}
\caption{Summation of background flow with the contribution of a close source.}
\label{sum}
\end{figure}

The mathematical model of expected flow was constructed as a sum of a close source's contribution and a galactic CR background (fig. \ref{sum}).

\begin{equation}
F_{sum} = F_{bgr}(R) + F_{star}(R)
\end{equation}

	
A simple power law spectrum $F = a_0 R^{-\gamma}$ was selected as galactic background, with parameters $a_0$ and $\gamma$ for each of the CR nuclei corresponding to spectra parameters measured by ATIC \citep{atic} and NUCLEON \citep{nucl} experiments in $50-3000$ GV magnetic rigidity region.


The contribution of a close source is calculated through a diffusion equation for a flash lamp like source (instantaneous and point-like). This approximation describes spatial distribution of supernova remnants class of sources well enough because distance to these sources is much larger than their size. Also this approximation works well for rather old CR sources (more than 10 000 years), but for younger sources it will provide only a qualitatively correct view. As will be shown, the last case (younger sources) will turn out to be more likely, therefore the flash lamp like approximation describes this article's problem only qualitatively and this article should be considered the first step on the way to solving the close source problem.


The source spectra is described by a double power law function with a kink and a smoothed out transition between them \citep{horandel}:

\begin{equation}
dQ(R) = R^{-\gamma_0} \left ( 1 + \left (\frac{R}{R_{ref}}\right )^{\omega_0} \right )^{-\Delta\gamma_0/\omega_0} dR,
\label{src_eq}
\end{equation}


\noindent where $R$ is magnetic rigidity, $R_{kink}$ -- kink rigidity, $\gamma_0$ -- spectre inclination before the kink, $\Delta\gamma_0$ -- difference in inclinations before and after the kink, $\omega_0$ -- smoothening coefficient.

\begin{figure}[!t]
\begin{center}
\includegraphics*[width=0.9\textwidth]{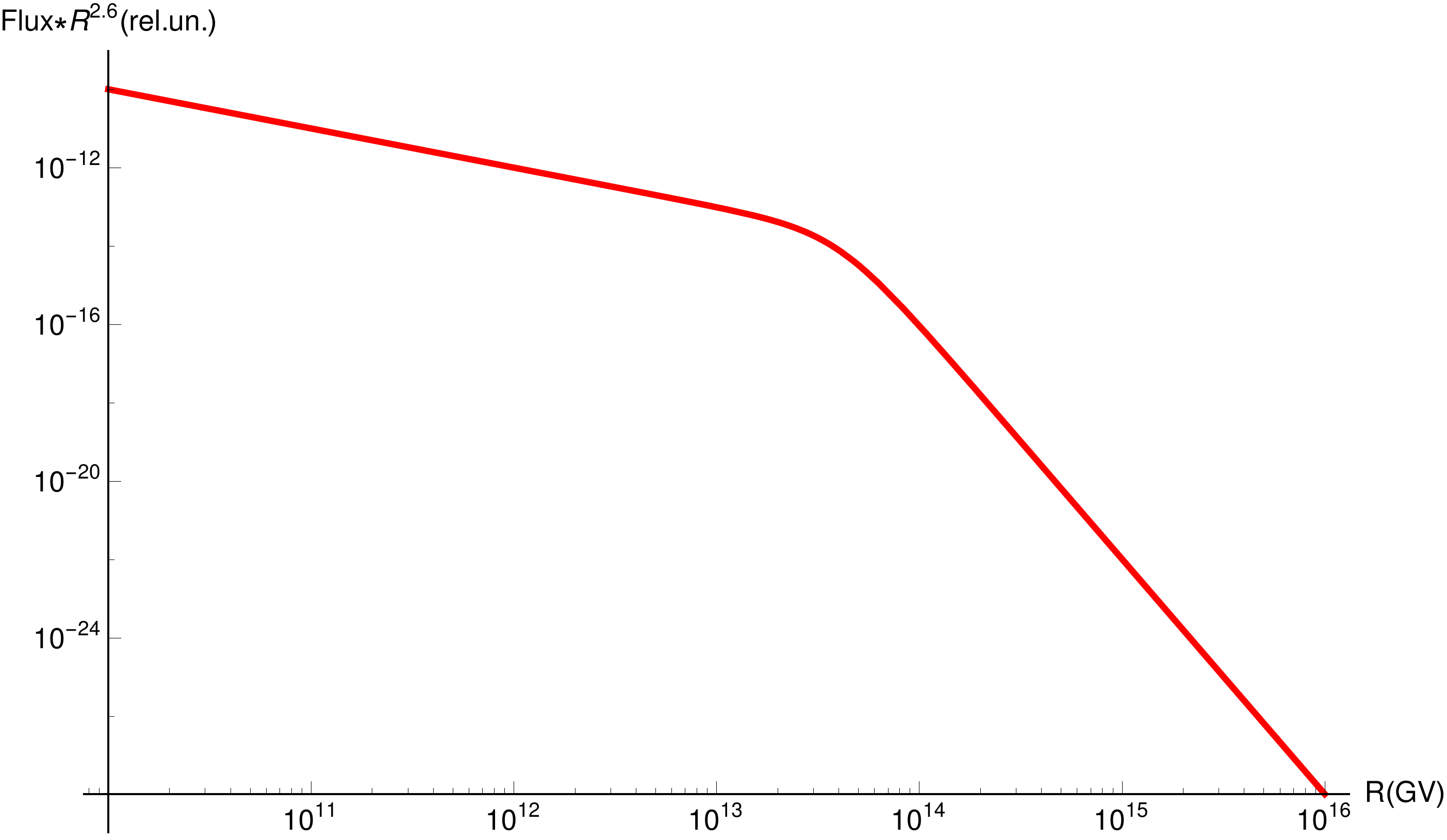}
\end{center}
\caption{Source spectre.}
\label{srcsp}
\end{figure}

In this model source spectra for different CR components are identical, only their integral intensity differs (fig. \ref{srcsp}).

Total energy of CR generated by the source is

\begin{equation}
W = \int_{10^9}^{\infty} \left [ R^{-\gamma_0} \left ( 1 + \left (\frac{R}{R_{kink}}\right )^{\omega_0} \right )^{-\Delta\gamma_0/\omega_0} \right ] dR \sum_{i=1}^{28} \left ( Z_i Ab_i \right ),
\end{equation}


\noindent where $Z_i$ is charge of a nuclei, $Ab_i$ -- its relative intensity in the CR flow generated by the source. By different estimations \citep{svesh} energy spent by the source on CR acceleration is less than 10 \% of its total energy. 


The diffusion equation for a close source is

\begin{equation}
\frac{\partial N(R)}{\partial t} - \nabla ( D(R) \nabla N ) = Q(R,t,r),
\end{equation}


\noindent where $N$ is CR concentration, $Q$ -- source function, $D(R)$ -- diffusion coefficient calculated from

\begin{equation}
D_{xx}(R) = D_{xx0} \left ( \frac{R}{R_{ref}} \right )^{\delta},
\label{diff_eq}
\end{equation}


\noindent where $D_{xx0}$, $R_{ref}$ and $\delta$ are used from \citep{mcmc}.


Because flow of an instantaneous point-like source is just a Green's function of the diffusion equation, the CR flux $F$ satisfying the equation (\ref{diff_eq}) for a point-like instantaneous source with spectre shown in (\ref{src_eq}) is described by

\begin{equation}
F_z(R,r,t) = \frac{C}{4\pi}G(R,r,t)Q_z(R),
\end{equation}


where $R$ is rigidity, $r$ -- distance to the source, $t$ -- its age, $G(R,r,t)$ -- Green's function for the equation (\ref{diff_eq}). The simplest kind of Green's function for three dimentional diffusion in endless space was used:

\begin{equation}
G = \left ( \frac{1}{4 \pi k t} \right )^{3/2}e^{-r^2/4kt}
\end{equation}



\begin{figure}[!t]
\begin{center}
\includegraphics*[width=0.9\textwidth]{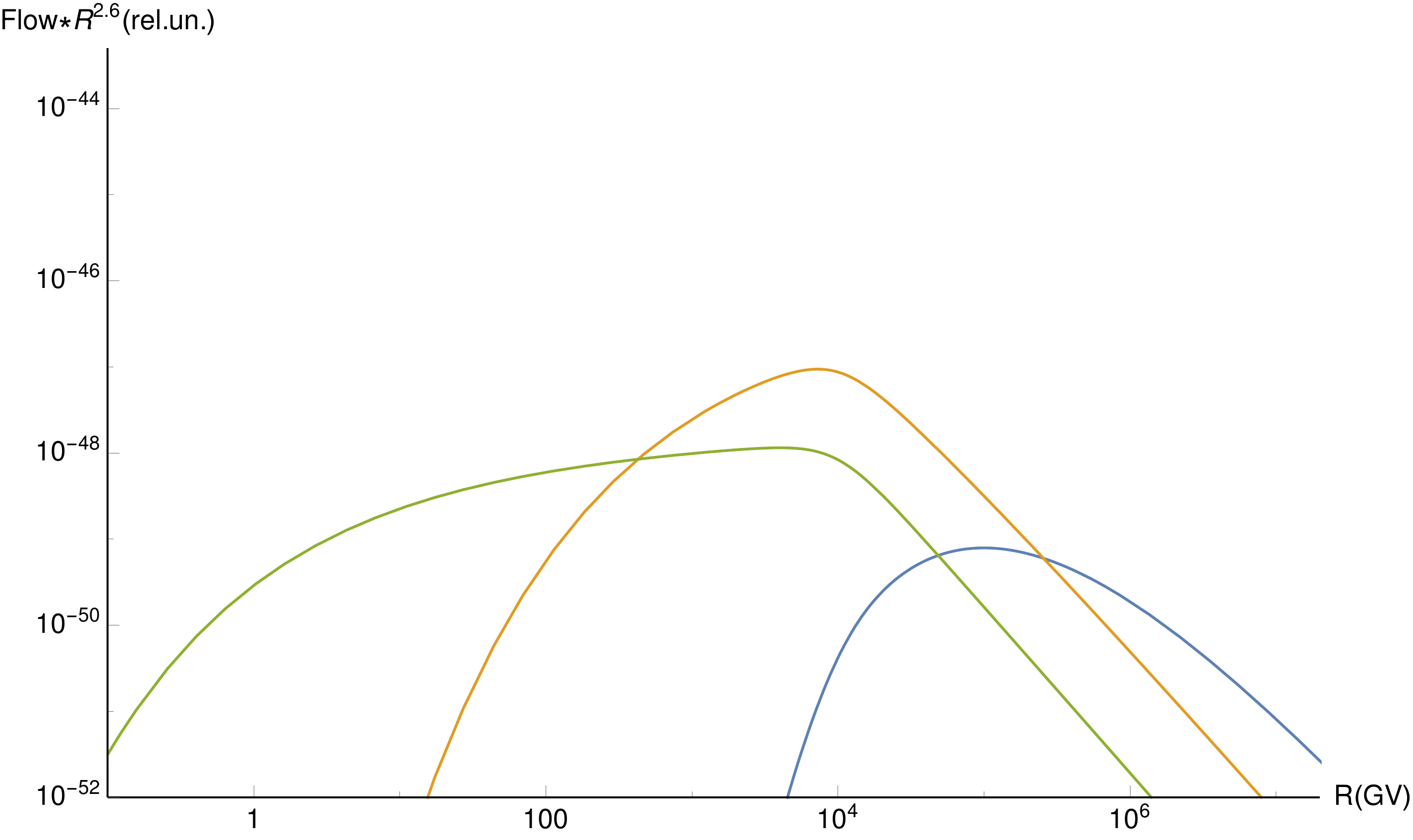}
\end{center}
\caption{Predicted shape of the contribution of a point-like source on a distance of 0.1 kPc from the observer for different ages: blue is for 0.1 kyear, orange -- 1 kyear, green -- 10 kyears.}
\label{agedep}
\end{figure}

Because diffusion coefficient depends on magnetic rigidity, flux from a source with spectre given in equation (\ref{src_eq}) will depend on its age and distance (fig. \ref{agedep}).

\section{Experimental data}


As mentioned earlier, background spectra for 50 GV -- 3000 GV rigidities were obtained from the ATIC balloon experiment \citep{atic} and from the NUCLEON experiment \citep{nucl}. Experimental data in the ``small knee'' region were obtained by the NUCLEON experiment which utilizes two independent energy measurement methods: classic ionization calorimeter method and the kinematic method KLEM \citep{nim}. Elemental spectra obtained by the NUCLEON experiment show an irregularity near 10 TV - hardening before this value and abrupt softening after.


Based on this data we selected 10 TV as the $R_{kink}$ parameter value and did not change that value throughout this work. 


Despite the fact that the NUCLEON experiment measures spectra of CR components from protons to Nickel nuclei, only protons, helium, carbon and oxygen spectra have high enough statistics in the ``small knee'' region, i.e. in the 10 TV region and higher, and only spectra of these nuclei were used to search for optimal parameters of the source. 

\begin{figure}[!t]
\begin{center}
\includegraphics*[width=0.9\textwidth]{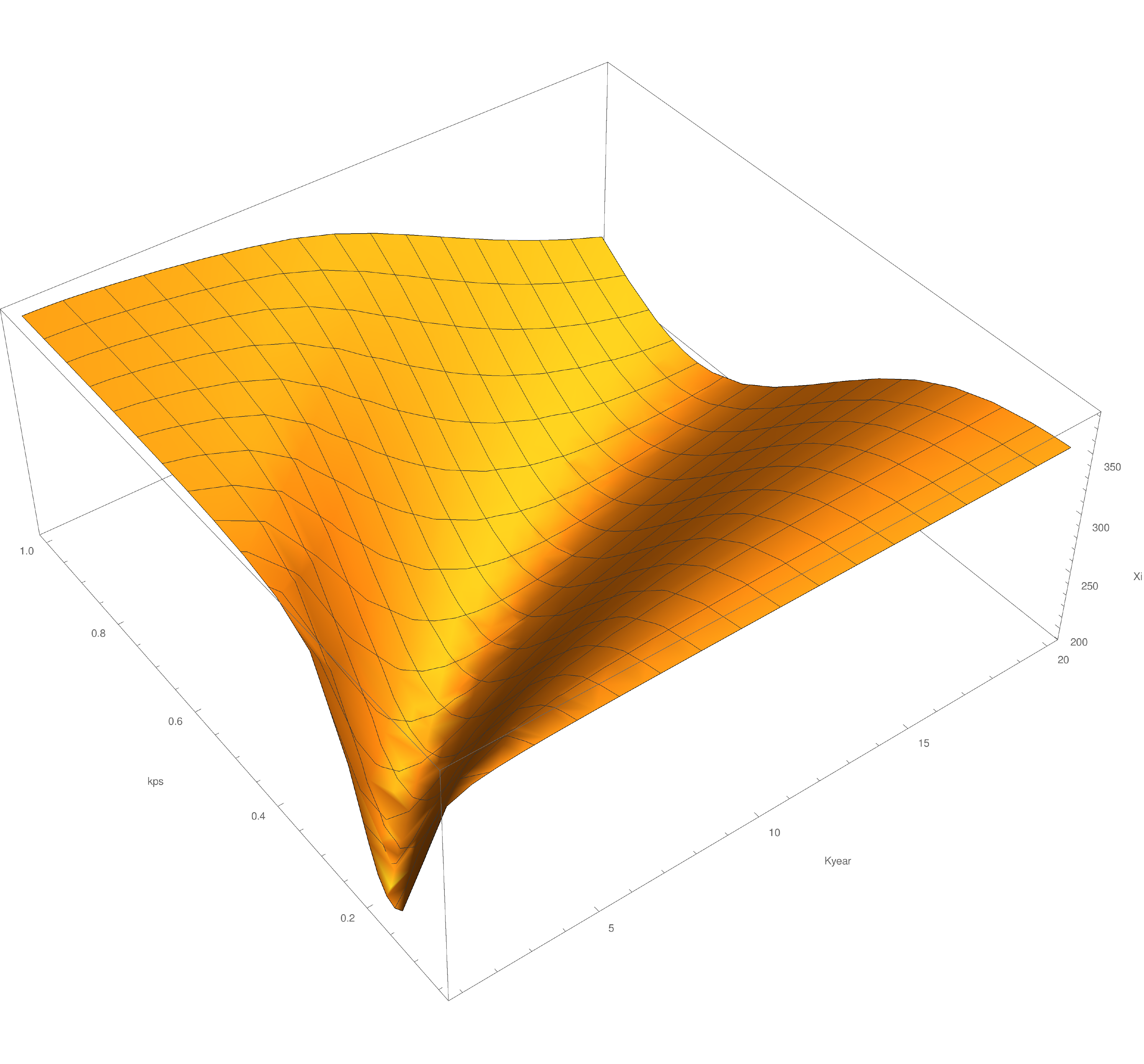}
\end{center}
\caption{Minimal $\chi^2$ value dependence on distance and age.}
\label{chi2}
\end{figure}

\section{Approximation}


To search for a hypothetical close source which can describe the mentioned irregularity of the spectra, we need to find suitable parameters for its distance, age, explosion energy (under the assumption that the CR source is a supernova remnant) and abundances of protons, helium, carbon and oxygen nuclei. To achieve this, for every point in age and distance space $(t,r)$ the following functional was minimized:

\begin{figure}[!t]
\begin{center}
\includegraphics*[width=0.9\textwidth]{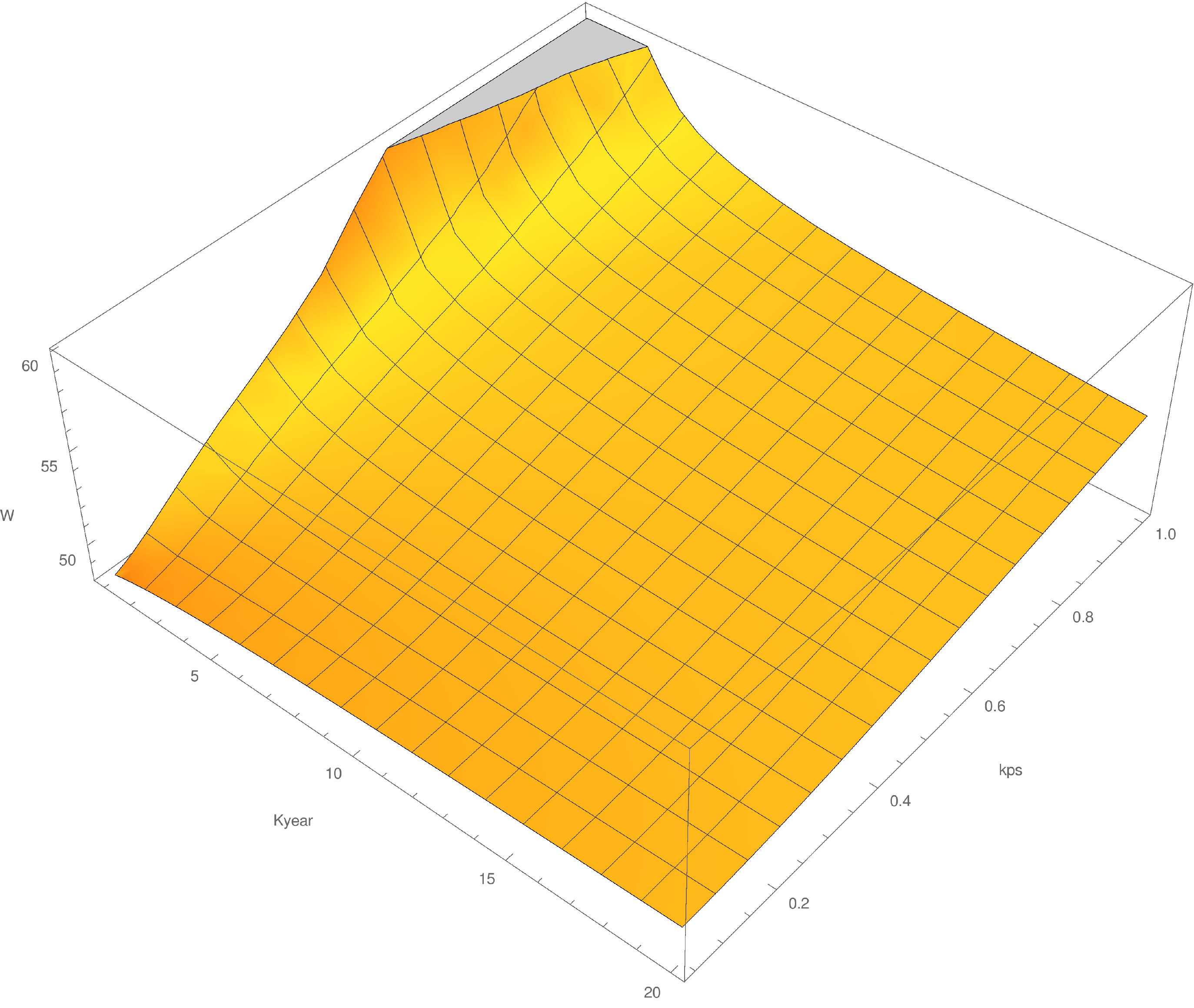}
\end{center}
\caption{$lg[W(t,r)]$ dependence on distance and age.}
\label{w}
\end{figure}

\begin{equation}
\chi^2 = \sum_{Z=1,2,6,8} \sum_i \left \{ \frac{F_i^Z - M_i^Z(T,r,W,Ab_{He},Ab_C,Ab_O)}{\sigma_i^Z} \right \}^2,
\label{chi2_eq}
\end{equation}


where summation is done for all available experimental points $F_i^Z$ of proton, helium, carbon and oxygen spectra, $M_i^Z$ is the predicted flow of the CR for charge $Z$ and rigidity bin $i$, $\sigma_i^Z$ -- statistical error of the experimental measurement for charge $Z$ and rigidity bin $i$, $W$ -- total energy of the source calculated as a sum of proton, helium, carbon and oxygen energies, $Ab_{He}$, $Ab_C$ and $Ab_O$ -- integral spectra of corresponding nuclei relative to protons (proton relative integral intensity is 1).


\begin{figure}[!t]
\begin{center}
\includegraphics*[width=0.9\textwidth]{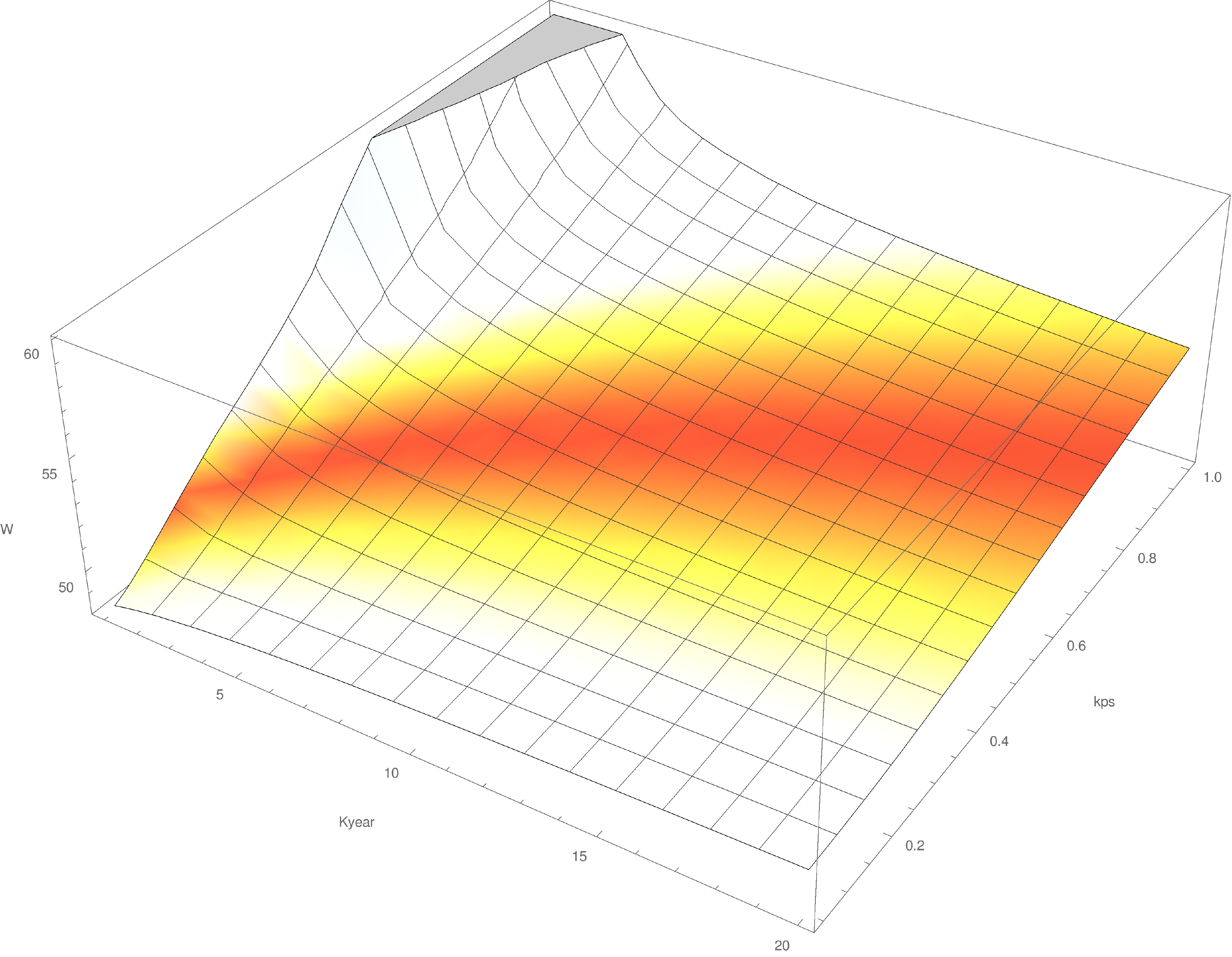}
\end{center}
\caption{$lg[W(t,r)]$ surface with corresponding $\chi^2$ values indicated in colour.}
\label{wcolor}
\end{figure}

There are 45 experimental points in a function with 4 free parameters for each pair of the $(t,r)$ parameters, therefore number of degrees of freedom $ndf = 41$.

\section{Results}


For each point of $(t,r)$ space value of $\chi^2$ (\ref{chi2_eq}) was minimized with $W$, $Ab_{He}$, $Ab_C$ and $Ab_O$ parameters. The obtained surface $\chi^2(t,r)$ has a pronounced region of minimal values (fig. \ref{chi2}).


$W(t,r)$ values (fig. \ref{w}) monotonously increase with $r$ which is in good agreement with predictions of a simplistic model (no diffusion, linear propagation), where contribution from a point-like source decreases as $1/r^3$. 


\begin{figure}[!t]
\begin{center}
\includegraphics*[width=0.9\textwidth]{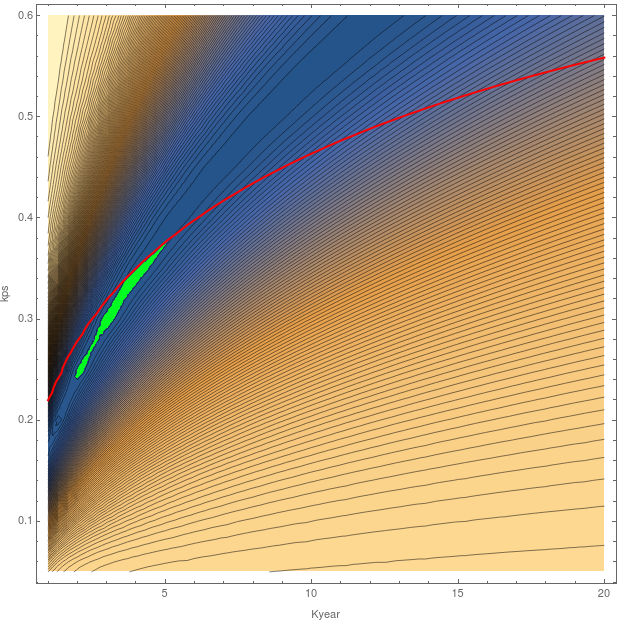}
\end{center}
\caption{A contour map where $\chi^2$ does not change along the contour lines and its value is presented as colour (blue corresponds to $\chi^2 = 190$, yellow -- $\chi^2 = 320$). The red line corresponds to maximum acceptable energy of the source ($W < 10^{51}$ erg).}
\label{contour}
\end{figure}

An abrupt increase in the necessary energy of the source in the low $t$ and high $r$ region is caused by the fact that CR of these energies do not have time to reach the observation point. We consider acceptable total energy of the source $W < 10^{51}$ erg, because by modern notion supernova explosion energy does not exceed this value \citep{svesh}.


On fig. \ref{wcolor} the $W(t,r)$ surface is coloured with values of corresponding $\chi^2$. Fig. \ref{contour} shows a contour map where $\chi^2$ does not change along the contour lines and its value is presented as colour. The red line corresponds to maximum acceptable energy of the source.



\begin{figure}[!t]
\begin{center}
\includegraphics*[width=0.9\textwidth]{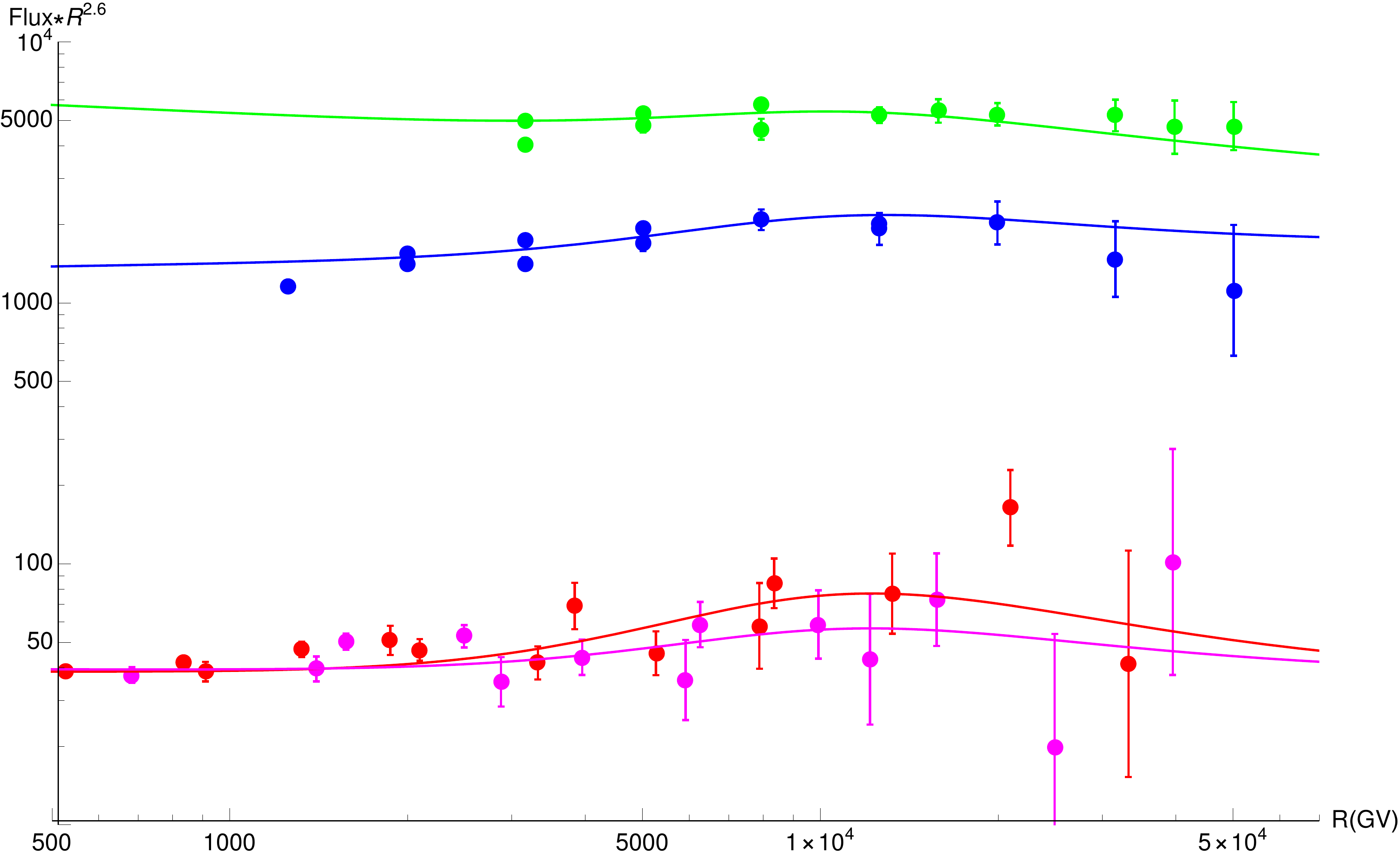}
\end{center}
\caption{Proton (green), helium (blue), carbon (red) and oxygen (magenta) spectra for the optimal source.}
\label{spectra}
\end{figure}
\begin{figure}[!t]
\begin{center}
\begin{minipage}{0.45\textwidth}
\begin{center}
\includegraphics*[width=0.9\textwidth]{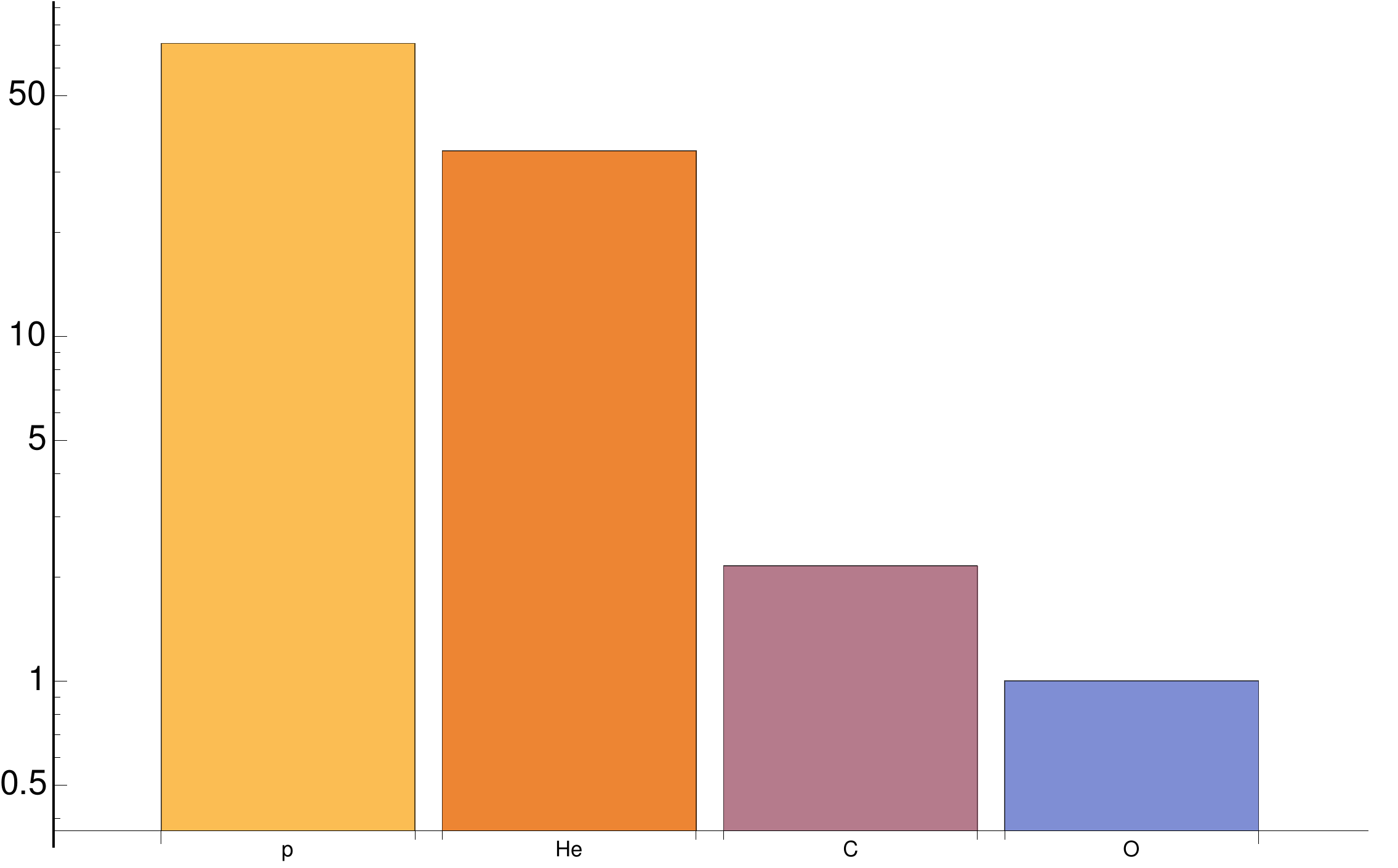}
\end{center}
\end{minipage}
\begin{minipage}{0.45\textwidth}
\begin{center}
\includegraphics*[width=0.9\textwidth]{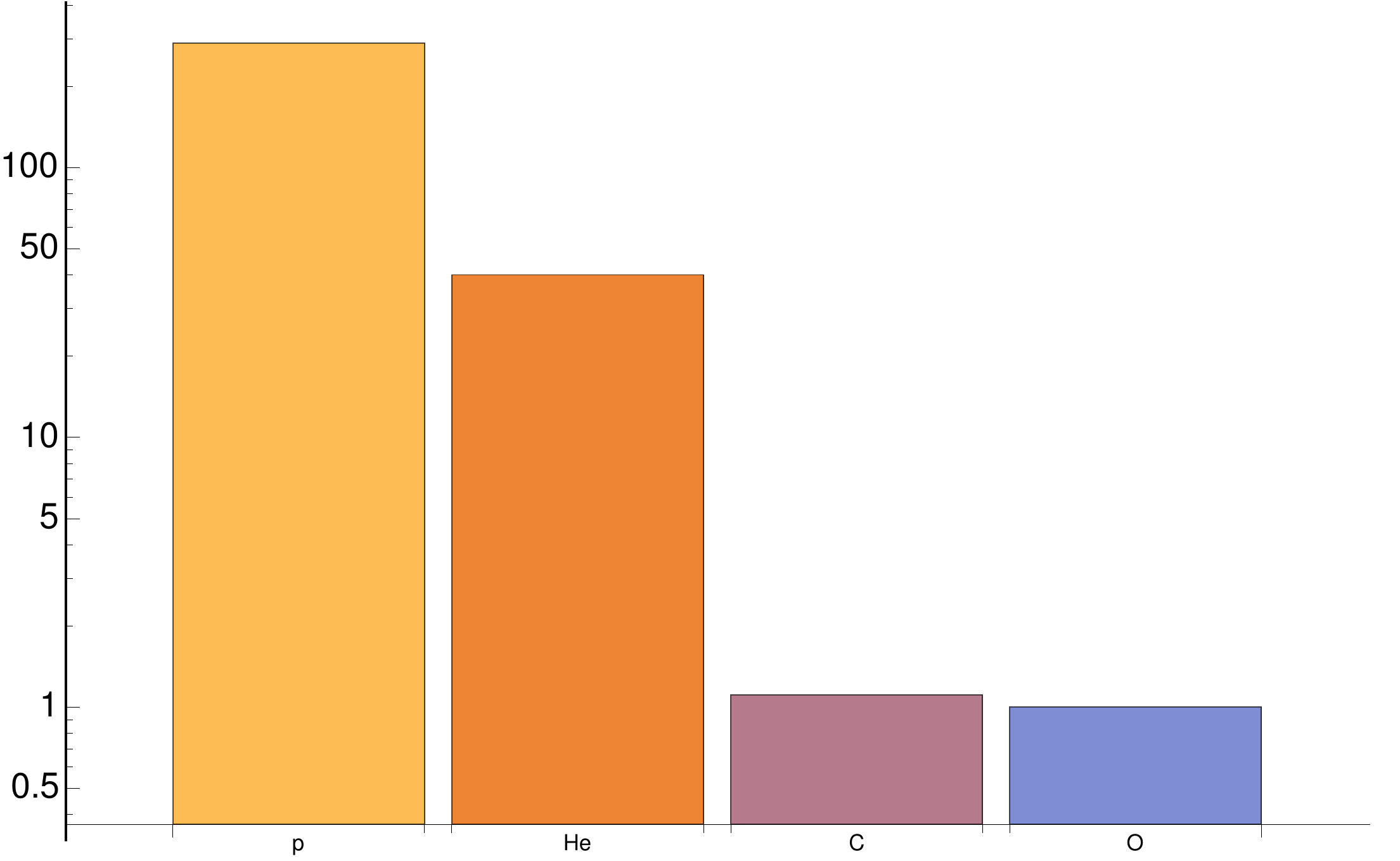}
\end{center}
\end{minipage}
\end{center}
\caption{Comparison of chemical composition of the optimal source (left) and galactic CR (right).}
\label{relch}
\end{figure}

Optimal position of the source corresponds to minimal value of the functional ($\chi^2 < 190/41$) within the energy constraint (marked green on fig. \ref{contour}, age from 2 to 5 kyears, distance from 0.22 to 0.36 kPc). Spectra expected from this hypothetical source are shown in fig. \ref{spectra}. Chemical composition of the source is presented in fig. \ref{relch}.

There are several known sources which could explain the irregularity of the spectra: 
\begin{itemize}
\item Vela \citep{vela} ($\chi^2=298/41$) with distance 0.25 -- 0.3 kPc, age 9 -- 27 kyears;
\item Lupus Loop \citep{loop} ($\chi^2=230/41$) with distance 0.15 -- 0.5 kPc, age 15 -- 31 kyears;
\item HB9 \citep{hb9} ($\chi^2=203/41$) with distance 0.4 -- 1.2 kPc, age 4 -- 7 kyears.
\end{itemize}

\section{Conclusions}

In this article a model of a single close flash lamp like point-like source contribution to the background CR spectra with no energy losses or fragmentation during propagation is presented. This model is in a reasonable agreement with experimental data and it illustrates a possibility of explanation of the observed CR ``small knee'' near 10 TV rigidity in terms of a single close source with total energy less than $10^{51}$ erg, out of which only 10 \% goes into accelerating charged components of CR.


The model predicts the most probable distance to such a hypothetical source of 0.22 -- 0.36 kPc and an age of 2 to 5 kyears.


Some of the observed sources such as Vela, Lupus Loop and HB9 are reasonable candidates for such a source, but their agreement with experimental data is worse than for the hypothetical one.

\end{document}